\documentclass{appolb}
\usepackage{graphicx}
\usepackage{amsmath}
\usepackage{psfrag}
\usepackage{latexsym}

\newcommand{\bq}{\begin{equation}}
\newcommand{\eq}{\end{equation}}
\newcommand{\ba}{\begin{eqnarray}}
\newcommand{\ear}{\end{eqnarray}}
\newcommand{\1}{\mathbf{1}}
\newcommand{\X}{\mathbf{X}}

\newcommand{\ec}{\mathbf{c}}
\newcommand{\C}{\mathbf{C}}

\newcommand{\A}{\mathbf{A}}

\newcommand{\rr}{\vec{r}}
\newcommand{\F}{\vec{F}}

\begin{document}
\title{Eigenvalue density of empirical covariance matrix for
correlated samples}
\author{Zdzis\l{}aw Burda, Jerzy Jurkiewicz, Bart\l{}omiej Wac\l{}aw
\address{Mark Kac Center for Complex Systems Research and
Marian Smoluchowski Institute of Physics,
Jagellonian University, Reymonta 4, 30-059 Krak\'ow, Poland}
}

\maketitle

\begin{abstract}{ 
We describe a method to determine
the eigenvalue density of empirical covariance
matrix in the presence of correlations between samples. 
This is a straightforward generalization of 
the method developed earlier by the authors for uncorrelated 
samples \cite{bgjw}. The method allows for exact 
determination of the experimental spectrum for a given
covariance matrix and given correlations between samples
in the limit $N\rightarrow \infty$ and $N/T=r=\mbox{const}$
with $N$ being the number of degrees of freedom and $T$ being
the number of samples. We discuss the effect of correlations
on several examples. }
\end{abstract}

\bigskip

Spectral properties of empirical covariance matrices
have been intensively studied for uncorrelated samples \cite{mp,bs,cs}.
The eigenvalue distribution of the empirical covariance matrix
is related to the eigenvalue distribution of the true
covariance matrix by Mar\v{c}enko-Pastur equation \cite{mp}.
This relation plays an important role in many practical
problems ranging from physics, telecommunication \cite{m}
and information theory \cite{s} to biology and quantitative 
finance \cite{bclp}.
Recently the corresponding equations has also been derived
for correlated samples \cite{bjw}. 

We will discuss some practical applications of these equations.
More precisely we will show how to use them to explicitly
compute the eigenvalue density of the empirical covariance matrix when
the true covariance matrix is given.
This method is of importance for instance for the problem
of optimal portfolio selection if one considers strategies 
based on the analysis of very short time price changes
which are known to be correlated in time, or for problems
in telecommunication where one discusses propagation of
signal through a random medium from a certain number of senders
to a certain number of receivers (see e.g. \cite{sm} and references therein)
and for many other problems.

To set the stage let us briefly recall the mathematical
formulation of the problem. Let $\X=(X_{i\alpha})$ be a 
real rectangular matrix of dimension $N\times T$ representing 
sampled values of a statistical system with $N$-degrees of freedom 
obtained in $T$ measurements. $\alpha$-th column of the matrix 
$\X$ contains $N$ numbers obtained in the $\alpha$-th 
measurement.  The standard estimator of the covariance 
matrix $\C$, describing correlations the degrees of freedom, is given by the
matrix\footnote{We shall assume throughout the paper
that $\langle X_{i\alpha} \rangle =0.$}
$\ec=(1/T)\X\X^{\tau}$: $c_{ij} = (1/T) \sum_\alpha X_{i\alpha}X_{j\alpha}$
which is called empirical covariance matrix. $\X^\tau$ stands
for the transpose of $\X$. The Mar\v{c}enko and Pastur solution \cite{mp}
refers to the situation when the measurements are uncorrelated:
\bq
\langle X_{i\alpha} X_{j\beta} \rangle = \delta_{\alpha\beta} C_{ij}.
\eq
It relates the eigenvalue density of $\ec$ to that of $\C$.
In the presence of correlations   
the Kronecker delta is replaced by an arbitrary 
symmetric semi-positive matrix $\A$
describing correlations between measurements:
\bq
\langle X_{i\alpha} X_{j\beta} \rangle = A_{\alpha\beta} C_{ij}.
\label{2AC}
\eq
We shall discuss this case in the present paper. Our aim is to
recall the formal solution \cite{bjw} and to show how to
use it to calculate spectral density $\rho_\ec$ of the empirical covariance 
matrix $\ec$, for given matrices $\A,\C$.

The equations for the eigenvalue density $\rho_\ec$
can be derived by assuming that statistical
fluctuations in the studied system are Gaussian. 
Then the matrix $\X$ can be treated as a random
matrix chosen from the Wishart ensemble with the following 
probability measure \cite{bjw,w}:
\ba
P(\X) \; D\X = 
{\cal N} \exp\left[ -\frac{1}{2} \Tr \,
\X^\tau \C^{-1} \X \A^{-1} \right] \;
\prod_{i,\alpha=1}^{N,T} 
d X_{i\alpha} \ ,
\label{preal}
\ear
where the normalization constant is ${\cal N} =
(2\pi)^{-\frac{NT}{2}} (\det \C)^{-\frac{T}{2}} 
(\det \A)^{-\frac{N}{2}}$. One can easily check that
correlations between matrices chosen randomly from this
ensemble are given by (\ref{2AC}). Applying a diagrammatic
technique \cite{fz,ms,bgjj} one can calculate averages 
of various quantities,
in particular the resolvent of the empirical covariance matrix:
\bq
g_{\ec}(z) = \left\langle \rm{Tr} \frac{1}{z-\ec} \right\rangle 
= \left\langle \rm{Tr} \frac{1}{z-(1/T) \X \X^\tau} \right\rangle 
.
\eq
The brackets denote the average over the ensemble of 
random matrices $\X$ with the probability measure (\ref{preal}).
It is convenient to express $g_{\ec}$ in terms of generating functions
for spectral moments of $\C,\A$ and $\ec$:
\bq
M_\C (z) = \sum_{k=1}^\infty \frac{M_{\C k}}{z^k}, \;\; 
M_\A (z) = \sum_{k=1}^\infty \frac{M_{\A k}}{z^k}, \;\; 
m_\ec (z) = \sum_{k=1}^\infty \frac{m_{\ec k}}{z^k} ,
\label{MCmc}
\eq
where $M_{\C k} = \int \rho_\C (x) x^k dx$ are spectral
moments of the eigenvalue density of the matrix $\C$
and similarly $M_{\A}$ and $m_{\ec}$ of $\A$ and $\ec$. 
The generating functions
are related to the corresponding resolvents, in particular
$m_\ec(z) = z g_\ec(z) - 1$.

In the limit of $N\rightarrow \infty$ and fixed
``rectangularity'' coefficient $r=N/T$ 
the generating functions $M_\C$ and $m_\ec$ are related by
two equations \cite{bjw}:
\ba
&m_{\ec}(z)= M_{\C}(Z)  , \nonumber \\
&z=Z r M_{\C}(Z) M_{\A}^{-1}(r M_{\C}(Z)),
\label{mapAC}
\ear
where $M_{\A}^{-1}$ is the inverse function of the
generating function $M_{\A}$. From these equations one 
can formally calculate $m_{\ec}(z)$ when
$\A$ and $\C$ are given. The variable $Z$ is auxiliary.
Having determined $m_{\ec}(z)$ from the 
equations (\ref{mapAC}) one can calculate the
resolvent $g_{\ec}(z) = (m_{\ec}(z)+1)/z$ and the eigenvalue
density $\rho_{\ec}(z)$ of the empirical covariance matrix $\ec$:
\begin{eqnarray}
\rho_{\ec}(x) = -\frac{1}{\pi} \mbox{Im} \, g_{\ec}(x+i0^+) =
-\frac{1}{\pi} \mbox{Im} \, \frac{1 + m_{\ec}(x+i0^+)}{x+i0^+} \ .
\label{rhoc}
\end{eqnarray}
Before we show how to calculate $m_{\ec}(z)$ in the most general
case we will first recall the case without
correlations that is for $A_{\alpha\beta}$ equal to
the Kronecker delta. The equations (\ref{mapAC})
simplify to \cite{bgjj}:
\ba
&m_\ec(z) = M_\C(Z), \nonumber \\
&z=Z(1+r M_\C (Z)).
\label{map}
\ear
This set of equations is equivalent to the Mar\v{c}enko-Pastur
equation \cite{mp}. It can be directly used to
determine the eigenvalue density $\rho_{\ec}(x)$ \cite{bgjw}.
We will sketch the method below.
Let $\lambda_1<\dots<\lambda_k$ be ordered eigenvalues
of $C$ and $n_k$ their multiplicities. The generating function
$M_\C(Z)$ takes the form:
\bq
M_\C(Z) = \sum_{k=1}^K
\frac{p_k \lambda_k}{Z-\lambda_k},
\label{MCZ}
\eq
where $p_k = n_k/N$.
Using (\ref{map}) and (\ref{MCZ}) we can determine $m_\ec(z)$
and hence from Eq. (\ref{rhoc}) the eigenvalue density function $\rho_\ec(x)$.
The shape of the eigenvalue density $\rho_{\ec}(x)$
is encoded in the behavior of the conformal
map (\ref{map}) near the critical horizon \cite{bgjw},
defined as a curve on the $Z$-plane which is mapped by (\ref{map})
into an interval $[x_-,x_+]$ on the real axis in the $z$-plane.
Here $x_-,x_+$ denote the upper and the lower edge 
of the support of the function $\rho_\ec(x)$. The critical horizon 
may be determined as follows: let $Z=X+iY$ be a point on the horizon.
Because the eigenvalues are real the imaginary part of $z=z(Z)$ 
must vanish. This gives the equation:
\bq
\sum_k \frac{p_k\lambda_k^2}{(X-\lambda_k)^2+Y^2} = \frac{1}{r} .
\label{main}
\eq
Solving this equation for $Y=Y(X)$ for given $X$ we find 
the desired curve giving the critical horizon: $X+i Y(X)$.
The equation (\ref{main}) has two symmetric roots $\pm Y$ 
because of the ambiguity of the map $z=z(Z)$. 
They can be found numerically by an iterative method
taking as an initial value $Y_0$ a small positive number.
The positive root 
is mapped into $z=x+i0^+$. The corresponding values 
of $x=z(Z)$ and $\rho_\ec(x)$ are given by:
\bq
x(Z) = X + r \sum_k p_k \lambda_k + 
r \sum_k \frac{p_k\lambda_k^2(X-\lambda_k)}{
(X-\lambda_k)^2 + Y^2},
\eq
and
\bq
\rho_\ec(Z) = -\frac{Y}{\pi x(Z)} 
\sum_k \frac{p_k\lambda_k}{(X-\lambda_k)^2+Y^2} .
\label{eq:rhoc3}
\eq
Treating $Z$ as a dummy parameter 
we obtain a pair  $(x,\rho_\ec)$ which is equivalent to
the eigenvalue density $\rho_{\ec}(x)$.
Briefly speaking, the variable $Z=X+iY$ is used to parametrize both the
eigenvalue $x$ and the spectrum $\rho_\ec$. 

The variable $X$ is an independent variable in this
construction. It is restricted to a finite range
$X\in [X_-,X_+]$. The limits of this range 
are mapped into the lower, $x_-$,
and the upper edge, $x_+$, of the spectrum $\rho_{\ec}(x)$.
They can be determined numerically
by observing that they come from the points
where the critical horizon intersects the real axis. 
Setting $Y=0$ in Eq. (\ref{main}) we obtain an
equation for $X$ whose largest root corresponds to $X_+$ and
the smallest to $X_-$. The roots can be found numerically.

The advantage of this method in comparison with the others 
\cite{bs,cs} is that one has to solve only one algebraic 
equation for one point in the spectrum, which can be done either 
analytically (for $K\leq 4$ or some other special cases) or numerically.
A more detailed discussion can be found in \cite{bgjw}.

Encouraged by the success of this approach  
we want to adopt it to the general case of correlated 
samples (\ref{2AC}). We have to return to the 
equations (\ref{mapAC}). For given matrix $\C$ we can calculate
$M_\C(Z)$ (\ref{MCZ}) and similarly for $\A$:
\bq
  M_{\A}(Z') = \sum_{\alpha=1}^{\kappa} 
  \frac{p_\alpha \Lambda_\alpha}{Z'-\Lambda_\alpha}.
  \label{MAZ}
\eq
We have indexed the parameters of the spectrum of the matrix $\A$
by Greek indices to distinguish them from the parameters of the
spectrum of $\C$.
As before, the first step of the construction is to 
parametrize critical horizon $Z$ by one real variable. 
However, the situation is now more complicated because we have to
invert $M_\A$ which may give a multi-valued function.
Let us introduce the notation: $Z=X+iY$ and $M_{\A}^{-1} = U+iV$.
Now, $x(Z)=z(Z)$ takes the form:
\ba
 x = & & r\sum_k \frac{p_k\lambda_k \left[ U(X^2+Y^2)+\lambda_k(YV-XU)\right]}
 {(X-\lambda_k)^2+Y^2} + \nonumber \\
 & & + i r \sum_k \frac{p_k\lambda_k \left[ V(X^2+Y^2)-\lambda_k(XV+YU)\right]}
 {(X-\lambda_k)^2+Y^2},
 \label{zZ2}
\ear
and the condition $x\in R$ implies that
\bq
F_1(X,Y,U,V) \equiv 
\sum_k \frac{p_k\lambda_k \left[ V(X^2+Y^2)-\lambda_k(XV+YU)\right]}
{(X-\lambda_k)^2+Y^2} = 0 \label{f1}.
\eq
In order to invert $M_{\A}$ and to calculate $U,V$ we use the relation 
$r M_{\C}(X+iY) = M_{\A}(U+iV)$ which gives:
\ba
r \sum_k \frac{p_k\lambda_k(X-\lambda_k-iY)}{(X-\lambda_k)^2+Y^2}
= \sum_\alpha \frac{p_\alpha\Lambda_\alpha(U-\Lambda_\alpha-iV)}
{(U-\Lambda_\alpha)^2+V^2}.     \nonumber
\ear
Comparing the real and imaginary parts we get:
\ba
  F_2(X,Y,U,V) \equiv 
  r \sum_k \frac{p_k\lambda_k(X-\lambda_k)}{(X-\lambda_k)^2+Y^2}
  - \sum_\alpha \frac{p_\alpha\Lambda_\alpha(U-\Lambda_\alpha)}
  {(U-\Lambda_\alpha)^2+V^2} = 0, \label{f2} \ \ \ \\
  F_3(X,Y,U,V) \equiv r Y \sum_k \frac{p_k\lambda_k}{(X-\lambda_k)^2+Y^2}
  - V \sum_\alpha \frac{p_\alpha\Lambda_\alpha}{(U-\Lambda_\alpha)^2+V^2} 
  = 0 \label{f3} . \ \ \
\ear
For fixed $X$, the set of three equations (\ref{f1})-(\ref{f3}) has to be solved numerically
for three unknown variables $Y,U,V$.
Then we can use formulae (\ref{eq:rhoc3}) and (\ref{zZ2}) to
determine the spectrum $\rho_\ec(x)$, as it was done 
in the previous case for $\A=\1$.

These new equations are much more complicated than before 
and have no such a beautiful graphical
interpretation \cite{bgjw} as Eq. (\ref{main}). 
One can immediately realize that it is easy to eliminate 
the variable $U$ from Eq. (\ref{f1})
to obtain a set of two equations for two variables $Y,V$.
This makes however the equations less transparent and does not simplify
the calculations.

There are no universal root finding procedures
for a nonlinear set of equations. Moreover all of
them are usually very sensitive to the
choice of the initial condition and therefore it
is difficult to guarantee that the procedure will 
converge to the wanted solution starting from some initial
condition unless a special care is payed. In other words
we have to make some effort to fully automatize root finding 
on a desired Riemann sheet. The idea is to use
the continuity of the solution as we shall explain below.
Rewrite Eqs. (\ref{f1})-(\ref{f3}) 
in a compact form:
\bq
  \F(X,\rr) = 0, \label{ff}
\eq
where $\rr=(Y,U,V)=(r_1,r_2,r_3)$ and $\F=(F_1,F_2,F_3)$.
Instead of a single equation (\ref{main}) we have now 
three equations which must be solved for a 3-vector $\rr$.
The solution $\rr=\rr(X)$ is a continuous function of $X$.
For given $X$ different solutions $\rr(X)$ of Eq. (\ref{ff}) 
which lie on different Riemann sheets assume different 
values as long as they are outside the real axis. So they
can be distinguished by value.
Suppose that we know the value $\rr=\rr(X_0)$ for some $X_0$ and 
that $(Y(X_0),U(X_0),V(X_0))$ is the desired solution.
In order to calculate $\rr$ for $X=X_0+dX$ 
we can use the first term of the Taylor series:
\bq
  \rr(X_0+dX) \cong \rr(X_0) + 
  \left[ \frac{d \rr(X)}{d X} \right]_{X=X_0} \cdot dX
\eq
as an initial point for the root finding procedure.
The partial derivatives of $\rr$ with respect to $X$ can
be analytically found by differentiating (\ref{ff}) which gives:
\bq
\frac{d \rr(X)}{d X} = - \mathbf{Q} \; \frac{d \F}{d X} \ ,
\eq
where $\mathbf{Q}$ is the inverse of the matrix 
$(\partial F_i/\partial r_j)_{ij}$, whose elements
can be explicitly calculated from Eqs. (\ref{f1})-(\ref{f3}).
Because of the continuity of the solution
this procedure must be convergent to the solution
on the same Riemann sheet as for $X_0$
if $dX$ is small enough.
Then we can keep on repeating the whole procedure moving
in small steps $X\rightarrow X+dX$ along the same solution.
The procedure is stopped when $Y\leq 0$ because this means
that we have reached the point $\rho_\ec(x)=0$. 
Thus if we can guarantee that $\rr(X_0)$ is on the correct Riemann sheet,
all other $\rr(X)$ obtained in this procedure will be on the same sheet. 
The check, whether $\rr(X_0)$ lies on the desired Riemann sheet,
does not have to be very efficient since it is done once per run or
if the spectrum consists of disconnected parts, 
it should be done as many times as is the number of parts.

Let us remark that
the parameter $X$, being just the real part of $Z$,
which we used here to parametrize the
critical horizon is not always well suited to this purpose.
In general case one has to use a parameter which uniquely
parametrizes the solution. In all the cases discussed below except one,
$X$ does the job.

In some cases, where the map between $z$ and $Z$ is 
known explicitly one can simplify the method. 
Exponential correlations of samples, which are physically 
important, belong to this class:
\bq
A_{\alpha\beta} = \exp \left(-\frac{|\alpha -\beta|}{\tau} \right) .
\label{expdef}
\eq
In this case an exact formula
of the map $z=z(Z)$ is known\footnote{
Actually in \cite{bjw} rather a relation $Z=Z(z)$ is
given but it can be easily inverted for $z=z(Z)$.} \cite{bjw}:
\bq
z=Z\left( \coth(1/\tau)\cdot r M_\C(Z)+\sqrt{1+\frac{r^2}{\sinh(1/\tau)^2} M_{\C}^2(Z)} \right), 
\label{expz}
\eq
where $\tau$ 
is the range of correlations.
In the limit $\tau\rightarrow 0$, 
the quantities $\sinh(1/\tau)\rightarrow\infty,\; \coth(1/\tau)\rightarrow 1$ and
in consequence Eq. (\ref{expz}) simplifies to (\ref{map}) 
as it should. Because of the presence of the square root 
in (\ref{expz}) it is hard to divide the above formula 
into the real and imaginary part.
This is however not a problem for a numerical root finder.
Changing $X$ in small steps between the limiting values
$X_-$ and $X_+$ we solve the equation $\mbox{Im} \left(z( X+i Y )\right)=0$ 
for $Y=Y(X)$ and calculate $x=z(X+iY)$ and $\rho_\ec(x)$ by means 
of the formula (\ref{eq:rhoc3}).
The limiting values $X_\pm$ which correspond to the largest
and smallest real value on the horizon
can be determined by solving the equation
$\mbox{Im} \left(z( X+i Y )\right)=0$ 
for $X$ on real axis that is for $Y\rightarrow 0^+$. 

\begin{figure}
\psfrag{xx}{$X$} \psfrag{yy}{$Y$}
\begin{center}
\includegraphics[width=12cm]{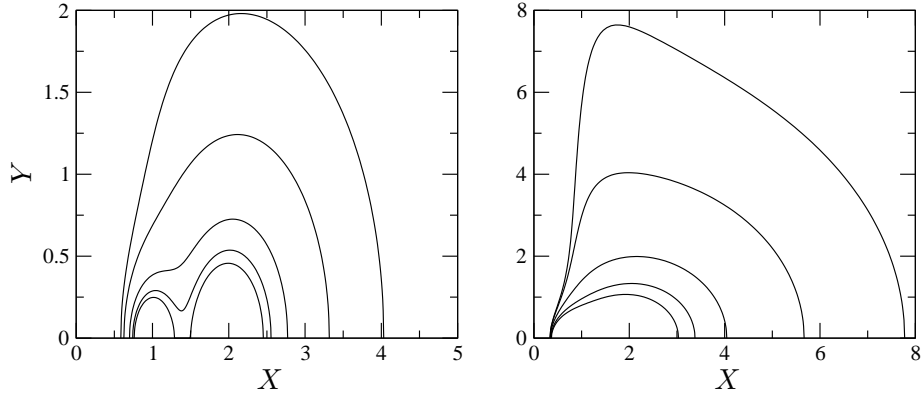}
\end{center}
\caption{Critical horizon for exponential 
correlations and $\{\lambda_i\}=\{1,2\}$.
The horizon is symmetric about the $X$ axis;
only upper part is shown. Left: $r=0.1$ and 
$\tau=0,1,2,5$ from the inner to the outer horizon.
Right: for $r=0.5$ and $\tau=0,1,2,5$.
Deviations from the shape 
expected in case of $\A=\1$ grow while the ratio 
$r/\sinh(1/\tau)$ increases.}
\label{exp+2ww}
\end{figure}

The influence of the correlation time $\tau$ on the map $z=z(Z)$ 
is illustrated in Fig. \ref{exp+2ww}.
Here $\C$ has two eigenvalues $\{\lambda_i\}=\{1,2\}$ with equal 
weights $p_1=p_2$.
One sees that for small $r$ and for $\tau$ being not very large 
the only effect of increasing $\tau$ is similar to increasing $r$.
However, for larger $r$ and growing correlation time
the map is deformed. This can be easily explained, because 
for small $r/\sinh(1/\tau) \ll 1$ the square root in the formula (\ref{expz})
is approximately equal 1. This means that 
formula (\ref{expz}) reduces
to the case (\ref{map}) without correlations but with new $r'=r\cdot \coth(1/\tau)$.
When the ratio $r/\sinh(1/\tau)$ becomes larger
the corrections begin to play an important role in
the map (\ref{expz}). 

\begin{figure}
\psfrag{xx}{$x$}\psfrag{yy}{$\rho_\ec(x)$}
\includegraphics[width=12cm]{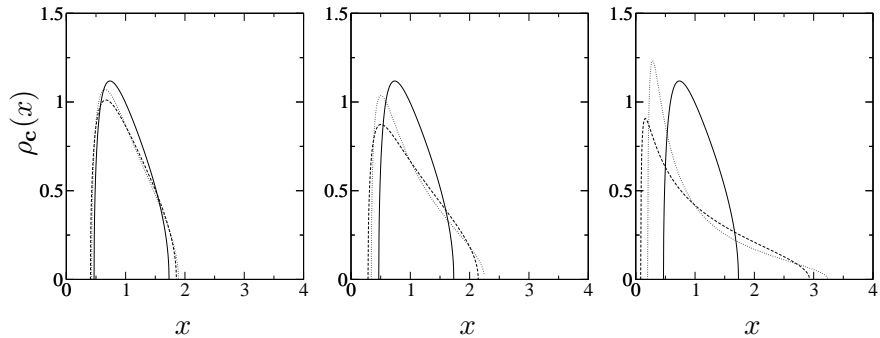}
\caption{Eigenvalues densities 
$\rho_\ec(x)$ for pure Wishart $\C=\A=\1$ with fixed $r=0.1$ 
(solid line), $\C=\1,r=0.1$ and exponential
correlations (\ref{expdef}) (dotted line) and
for pure Wishart with modified $r=0.1\cdot \coth(1/\tau)$ 
(dashed line) for different $\tau$: $\tau=1$ (left picture),
$\tau=2$ (middle picture), $\tau=5$ (right picture).
The distributions represented by dashed 
and dotted line behave similarly. In practice, when one
reconstructs them only from one set of
eigenvalues, they can be easily mixed up.
}
\label{exp-spec}
\end{figure}
This observation indicates that the presence of exponential correlations
between samples
may be confused with pure correlations ($\C\neq 1,\A=\1$) 
but with modified ``rectangularity'' coefficient $r$ (see Fig. \ref{exp-spec}).

\begin{figure}
\psfrag{xx}{$x$} \psfrag{yy}{$\rho_\ec(x)$}
\begin{center}
\includegraphics[width=10cm]{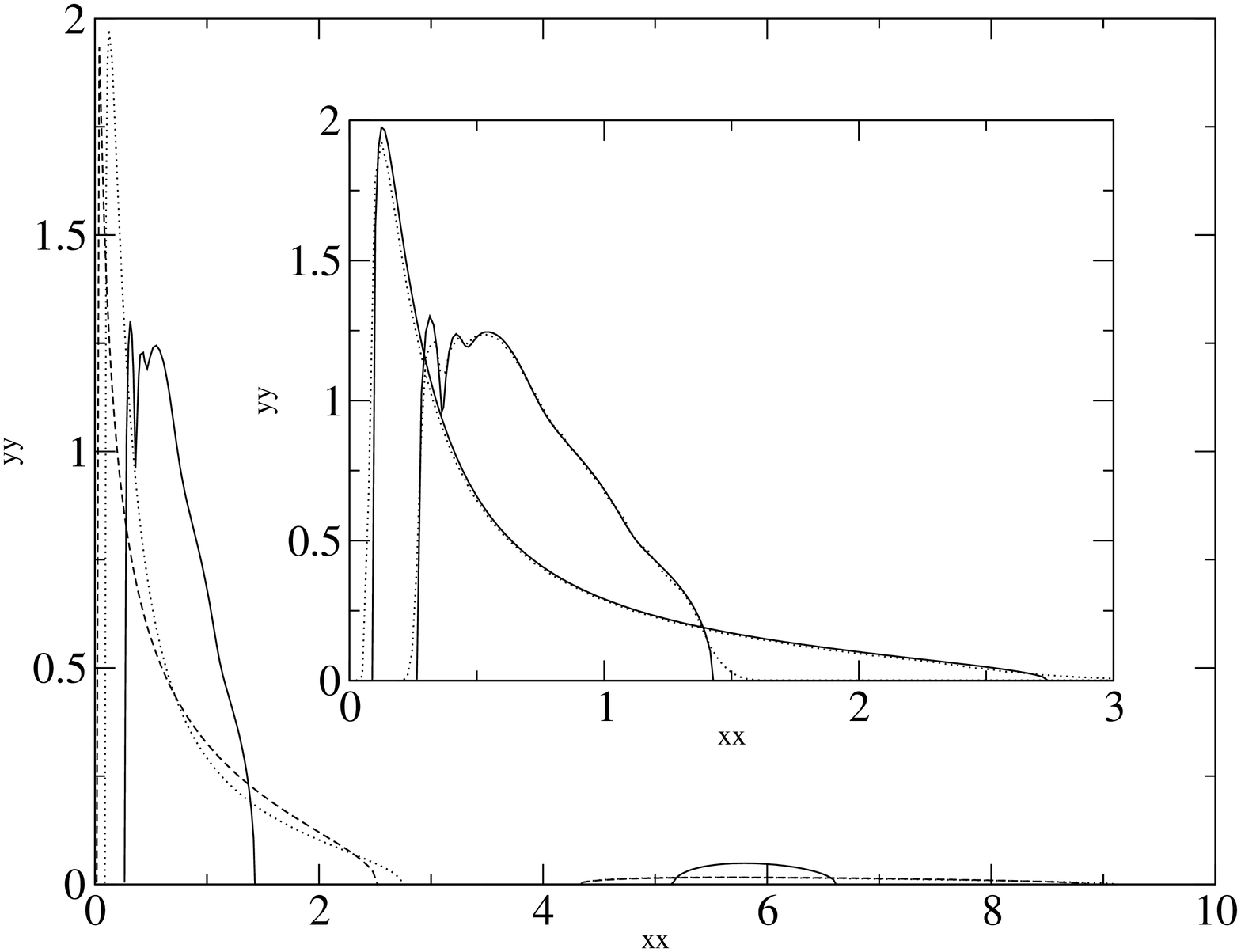}
\end{center}
\caption{The plot of $\rho_\ec(x)$ for $N=18$ eigenvalues of $\C$ taken from Polish Stock Market.
Solid: $r=18/255,\tau=0$, dotted: $r=18/255,\tau=10$, dashed: $r=180/255,\tau=0$.
Inset: the ``bulk'' of spectra for $r=18/255$ and $\tau=0,10$ calculated (solid line)
and found experimentally (dotted line) for sample of $3\times 10^5$ 
Wishart matrices of size $N=18$ generated by a Monte-Carlo procedure.}
\label{market}
\end{figure}
Consider now a practical example.
In the previous paper \cite{bgjw} we discussed the evolution
of the eigenvalue density of the experimental covariance matrix
$\ec$ with the number of independent measurements 
for $N=18$ eigenvalues, obtained from data for daily returns
of 18 stocks on the Polish Stock Market. Here we will discuss
how the spectrum changes in case of correlated returns as for example happens in short time horizons.
We will use the same eigenvalues as before \cite{bgjw}.
In the Fig. \ref{market}, $\rho_\ec$ is shown for the three cases:
for $r=18/255$ without correlations ($\tau=0$), for
$r=18/255$ and $\tau=10$, and finally for $r=10\cdot 18/255$ and $\tau=0$. 
We see that the first case deviates very much from the two remaining
ones. Comparing the second and the third case we see that they 
follow the same trends. Technically in the second example one has
ten times more data per degree of freedom but because the
correlation time is equal to ten this means that 
roughly only every tenth data point introduces
a new information. Therefore intuitively the second and 
the third case have a similar statistical content.
Indeed a quick look at Eq. (\ref{expz})
shows that if one approximates the contribution from the
square root by unity neglecting the term proportional
to $r^2$ then this equation will assume the same
form as Eq. (\ref{map}) with an effectively rescaled
parameter $r$: $r\rightarrow r \cdot \coth(1/10)\approx 10 r$. This
approximation is legitimate only if $r/\sinh(1/\tau) \ll 1$ 
as in the discussed example.

A similar observation can be made in a general case
of arbitrary correlations $\A$.
Consider the map (\ref{mapAC}) in case of small $r\ll 1$.
Assuming that the radius of the horizon is of order 
$\sqrt{r}$ as in case without correlations,
from (\ref{MCZ}) it follows 
that $|r M_\C(Z)|\approx \sqrt{r}$ on the horizon.
We can invert $M_\A(Z')$ for small $Z'$:
\bq
M^{-1}_{\A}(Z') = 
M_{\A 1} Z'^{-1} \left( 1 + \mu_1 Z' + \mu_2 Z'^2 + \dots\right),
\eq
where the coefficients of the series are expressed by the moments \cite{bjw}:
\bq
\mu_1 = \frac{M_{\A 2}}{(M_{\A 1})^2} \quad , \quad
\mu_2 = \frac{M_{\A 3} M_{\A 1} - 
(M_{\A 2})^2}{(M_{\A 1})^4} \quad , \quad \dots \quad .
\eq
Then the Eq. (\ref{mapAC}) takes the form:
\bq
z = M_{{\A}1} Z \left( 1 + \mu_1 r M_{\C}(Z) + 
\mu_2 r^2 M^2_{\C}(Z) + \dots \right).
\label{zZser}
\eq
The multiplicative factor $M_{\A 1}$  merely redefines $z$
and does not affect the map $z=z(Z)$. An important difference in comparison
with Eq. (\ref{map}) is the change of the 
coefficient at the linear term in $r$ which
tells us that for sufficiently small $r$ 
the map (\ref{zZser}) can be viewed as (\ref{map}) 
but with of a modified parameter $r$: 
$r\rightarrow r\cdot \mu_1 = rM_{\A 2}/(M_{\A 1})^2$.

To illustrate this, let us consider an example. Take 
$\C=\1$ and $\A$ having two 
eigenvalues one of which being $\Lambda_1=1$ and
the other $\Lambda_2$ being a free parameter.
Additionally assume that the two eigenvalues have
the same multiplicities and hence $p_1=p_2=1/2$. 
The model looks somewhat artificial but it 
well illustrates a feature which is quite general.

\begin{figure}
\psfrag{xx}{$\Lambda_2$}\psfrag{y1}{$X_-$}\psfrag{y2}{$x_-$}
\begin{center}
\includegraphics[width=12cm]{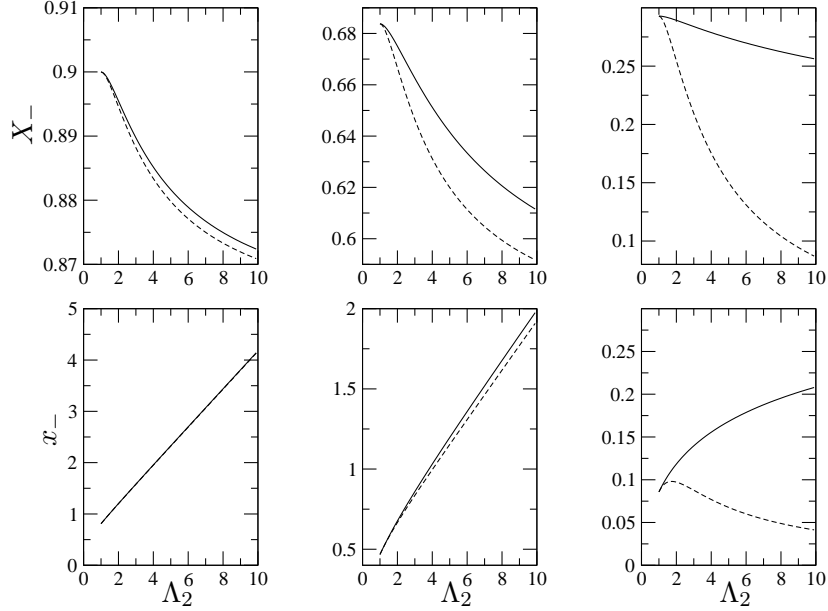}
\end{center}
\caption{Top: the position of left border $X_-$ of the map $z=z(Z)$
for $\C=\1$ and $\{\Lambda_\alpha\}=\{1,\Lambda_2\}$ (solid line)
and pure Wishart $\C=\A=\1$ with $r'=r\mu_1(\Lambda_2)$ (dashed line),
from the left: $r=0.01,0.1,0.5$. Bottom: 
the same for the left border $x_-$ of spectrum $\rho_\ec(x)$
with additionally rescaled $x$ axis: 
$x\rightarrow x \cdot M_{\A 1}$ for the Wishart case.
The deviations from the rescaled Wishart spectrum 
become significant for $r>0.1$.}
\label{2ww}
\end{figure}

In the upper part of Fig. \ref{2ww} we compare positions of 
the left border $X_-$ of the critical horizon (\ref{mapAC})
calculated in two different ways: numerically for different $r$,
and analytically for the Wishart model without correlations with
an effective ``rectangularity'' parameter $r' = \mu_1 \cdot r$,
which gives $X_-=1- \sqrt{r'}$ 
where $\mu_1=2(1+\Lambda_2^2)/(1+\Lambda_2)^2$.
The analogous comparison is made for the position of left
border $x_-$ of the eigenvalue density function $\rho_\ec(x)$
in the bottom of Fig. \ref{2ww}. Here we rescale additionally axes:
$x\rightarrow x \cdot M_{\A 1},y\rightarrow y/ M_{\A 1}$ with 
the factor $M_{\A 1}=(1+\Lambda_2)/2$ as in Eq. (\ref{zZser}).

\begin{figure}
\psfrag{xx}{$x$}\psfrag{yy}{$\rho_\ec(x)$}
\begin{center}
\includegraphics[width=10cm]{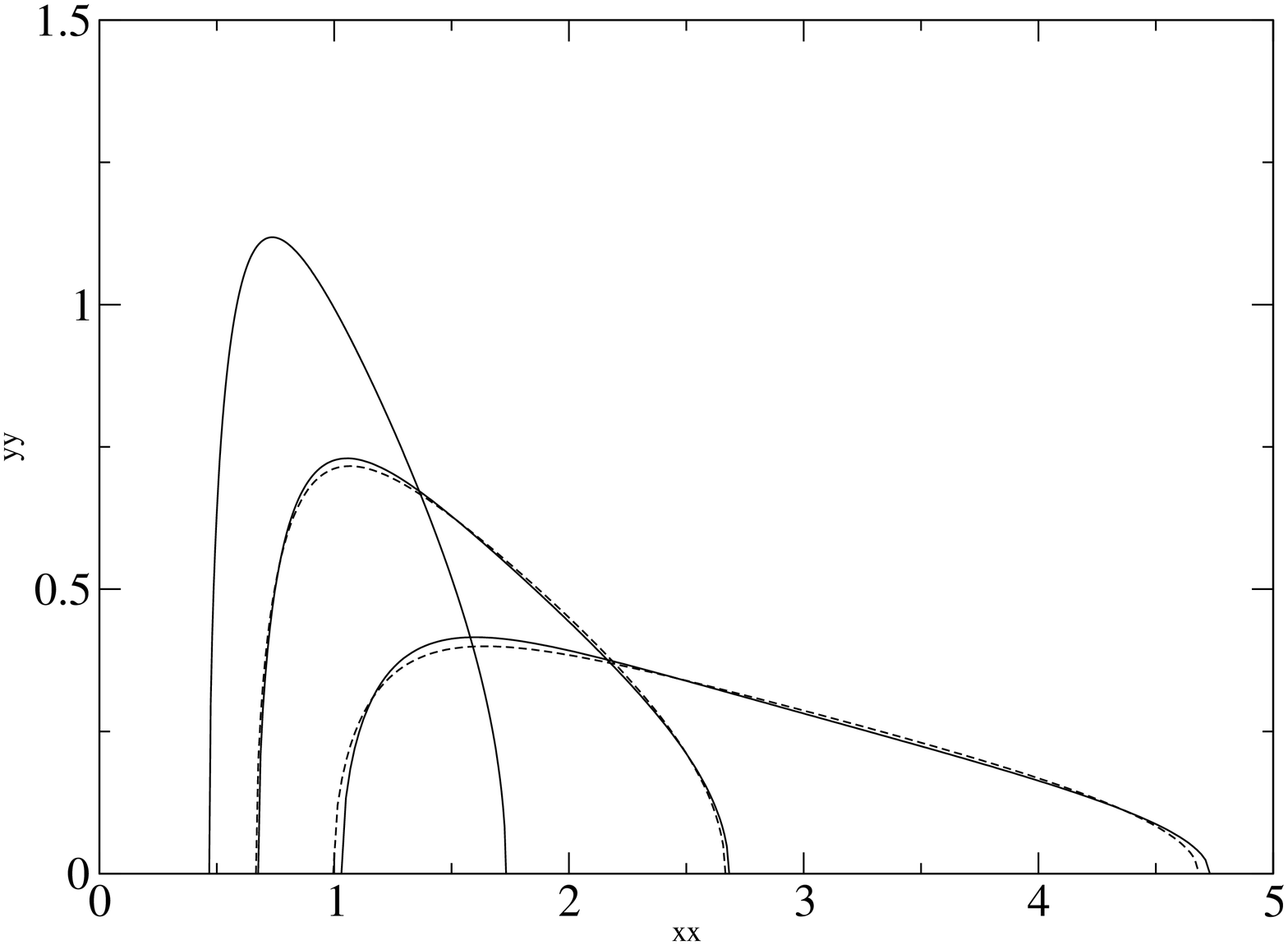}
\end{center}
\caption{The spectra for $r=0.1,\C=\1$ and $\A$ 
having two eigenvalues $\{1,\Lambda_2\}$ 
(solid line), and for rescaled Wishart (dashed line): 
$r\rightarrow r\mu_1,x\rightarrow x \cdot M_{\A 1},
y\rightarrow y/M_{\A 1}$. From the left to the
right: $\Lambda_2=1,2,4$.}
\label{2ww-rho}
\end{figure}

The corresponding eigenvalue distributions of the empirical
covariance matrix $\ec$ for these two cases for $r=0.1$
are shown in Fig. \ref{2ww-rho}. One sees an excellent agreement 
between them. This means that indeed 
for sufficiently small $r$ the only influence of $\A$ on the spectrum
of $\ec$ is an effective change of $r$ and rescaling the axes.

Let us shortly summarize. Using the relation between the 
moments generating functions (\ref{mapAC}) 
(or equivalently between the resolvents) 
we have shown how to
effectively compute the eigenvalue spectrum
of the empirically determined covariance matrix for given correlations,
even in the case of correlated
measurements. As an example of the application of the method
we have demonstrated the influence of the 
exponential correlations between measurements
on the eigenvalue spectrum of the covariance matrix calculated for
stocks' logarithmic returns. We have argued that
in the limit of the large number of samples that 
is for $T \gg N$, or equivalently for $r = N/T \ll 1$, the eigenvalue
density of the empirical covariance matrix can be approximated by
the eigenvalue density for a reduced number of uncorrelated samples, with the
reduction factor being approximately inversely proportional
to the correlation time for the original correlated 
samples.

\bigskip
\noindent
{\bf Acknowledgments}

This work was partially supported by
the Polish State Committee for Scientific Research (KBN) grant
2P03B-08225 (2003-2006) and Marie Curie Host Fellowship
HPMD-CT-2001-00108 and by EU IST Center of Excellence ``COPIRA''.

\end{document}